\documentclass[%
 reprint,
groupedaddress,
 amsmath,amssymb,
 aps,
]{revtex4-2}

\usepackage{graphicx}
\usepackage{dcolumn}
\usepackage{bm}

\newcommand{\GeV}{\textrm{GeV}}
\newcommand{\MeV}{\textrm{MeV}}

\begin{document}


\title{\large {\it White paper to the Proceedings of the
U.S.\ Particle Physics Community Planning Exercise (Snowmass 2021)}: \\
\vspace{0.0 cm}
\LARGE Linear colliders based on laser-plasma accelerators
\vspace{0.5 cm}}

\author{C. Benedetti}
\affiliation{Lawrence Berkeley National Laboratory, Berkeley, California 94720, USA}
\author{F. Albert}
\affiliation{{Lawrence Livermore National Laboratory, Livermore, California 94550, USA}}
\author{J. Bromage}
\affiliation{Laboratory for Laser Energetics, University of Rochester, Rochester, New York 14623, USA}
\author{D. Bruhwiler}
\affiliation{RadiaSoft LLC, Boulder, Colorado 80304, USA}
\author{S. S. Bulanov}
\affiliation{Lawrence Berkeley National Laboratory, Berkeley, California 94720, USA}
\author{E. M. Campbell}
\affiliation{Laboratory for Laser Energetics, University of Rochester, Rochester, New York 14623, USA}
\author{N. M. Cook}
\affiliation{RadiaSoft LLC, Boulder, Colorado 80304, USA}
\author{B. Cros}
\affiliation{Laboratoire de Physique des Gaz et des Plasmas, CNRS, Universit\'{e} 
Paris-Saclay, 91405 Orsay, France}
\author{M. C. Downer}
\affiliation{Department of Physics, University of Texas, Austin, Texas 78712, USA}
\author{E. Esarey}
\affiliation{Lawrence Berkeley National Laboratory, Berkeley, California 94720, USA}
\author{D. H. Froula}
\affiliation{Laboratory for Laser Energetics, University of Rochester, Rochester, New York 14623, USA}
\author{M. Fuchs}
\affiliation{Department of Physics and Astronomy, University of Nebraska, Lincoln, Nebraska 68588, USA}
\author{C. G. R. Geddes}
\affiliation{Lawrence Berkeley National Laboratory, Berkeley, California 94720, USA}
\author{S. J. Gessner}
\affiliation{SLAC National Accelerator Laboratory, Menlo Park, California 94025, USA}
\author{A. J. Gonsalves}
\affiliation{Lawrence Berkeley National Laboratory, Berkeley, California 94720, USA}
\author{M. J. Hogan}
\affiliation{SLAC National Accelerator Laboratory, Menlo Park, California 94025, USA}
\author{S. M. Hooker}
\affiliation{Department of Physics, Clarendon Laboratory, University of Oxford, Parks Road, Oxford OX1 3PU, UK}
\author{A. Huebl}
\affiliation{Lawrence Berkeley National Laboratory, Berkeley, California 94720, USA}
\author{C. Jing}
\affiliation{Euclid Techlabs LLC, Bolingbrook, Illinois 60440, USA}
\author{C. Joshi}
\affiliation{
University of California, Los Angeles, California 90095, USA}
\author{K. Krushelnick}
\affiliation{Center for Ultrafast Optical Science, University of Michigan, Ann Arbor, Michigan 48109, USA}
\author{W. P. Leemans}
\affiliation{Deutsches Elektronen-Synchrotron DESY, 22607 Hamburg, Germany}
\author{R. Lehe}
\affiliation{Lawrence Berkeley National Laboratory, Berkeley, California 94720, USA}
\author{A. R. Maier}
\affiliation{Deutsches Elektronen-Synchrotron DESY, 22607 Hamburg, Germany}
\author{H. M. Milchberg}
\affiliation{
University of Maryland, College Park, Maryland 20742, USA}
\author{W. B. Mori}
\affiliation{
University of California, Los Angeles, California 90095, USA}
\author{K. Nakamura}
\affiliation{Lawrence Berkeley National Laboratory, Berkeley, California 94720, USA}
\author{J. Osterhoff}
\affiliation{Deutsches Elektronen-Synchrotron DESY, 22607 Hamburg, Germany}
\author{J. P. Palastro}
\affiliation{Laboratory for Laser Energetics, University of Rochester, Rochester, New York 14623, USA}
\author{M. Palmer}
\affiliation{Accelerator Test Facility, Brookhaven National Laboratory, Upton, New York 11973, USA}
\author{J. G. Power}
\affiliation{Argonne National Laboratory, Lemont, Illinois 60439, USA}
\author{K. P\~{o}der}
\affiliation{Deutsches Elektronen-Synchrotron DESY, 22607 Hamburg, Germany}
\author{C. B. Schroeder}
\email[Corresponding author: ]{CBSchroeder@lbl.gov}
\affiliation{Lawrence Berkeley National Laboratory, Berkeley, California 94720, USA}
\author{B. A. Shadwick}
\affiliation{Department of Physics and Astronomy, University of Nebraska, Lincoln, Nebraska 68588, USA}
\author{D. Terzani}
\affiliation{Lawrence Berkeley National Laboratory, Berkeley, California 94720, USA}
\author{M. Th\'{e}venet}
\affiliation{Deutsches Elektronen-Synchrotron DESY, 22607 Hamburg, Germany}
\author{A. G. R. Thomas}
\affiliation{Center for Ultrafast Optical Science, University of Michigan, Ann Arbor, Michigan 48109, USA}
\author{N. Vafaei-Najafabadi}
\affiliation{Stony Brook University, Department of Physics and Astronomy, Stony Brook, New York 11794, USA}
\author{J. van Tilborg}
\affiliation{Lawrence Berkeley National Laboratory, Berkeley, California 94720, USA}
\author{M. Turner}
\affiliation{Lawrence Berkeley National Laboratory, Berkeley, California 94720, USA}
\author{J.-L. Vay}
\affiliation{Lawrence Berkeley National Laboratory, Berkeley, California 94720, USA}
\author{T. Zhou}
\affiliation{Lawrence Berkeley National Laboratory, Berkeley, California 94720, USA}
\author{J. Zuegel}
\affiliation{Laboratory for Laser Energetics, University of Rochester, Rochester, New York 14623, USA}

\date{March 15, 2022}

\maketitle


\begin{widetext}
\section*{Executive Summary}

Laser-plasma accelerators (LPAs) are capable of sustaining accelerating fields of 1--100 GeV/m, 10--1000 times that of conventional RF technology, and  the highest fields produced by any of the widely researched advanced accelerator concepts. Furthermore, LPAs intrinsically produce short particle bunches, 100--1000 times shorter than that of conventional RF technology, which leads to reductions in beamstrahlung and savings in  overall power consumption.  Furthermore, they enable novel energy recovering methods that can reduce power consumption and improve the luminosity per unit energy consumption for linear colliders. These properties make LPAs a promising candidate as drivers for a more compact, less expensive high-energy collider by providing  multi-TeV polarized leptons in a relatively short distance $\sim 1$ km. Collider concepts are discussed up to the 15 TeV range. A future RF-based linear collider facility could be re-purposed to delivery higher energies with LPA technology thereby extending physics reach while saving on construction costs.

Previous reports have made recommendations for a vigorous program on LPA R{\&}D and applications, including the previous P5 and subcommittee reports, the European Strategy and Laboratory Directors Group reports, and several others. Numerous significant results have been obtained since the last P5 report, including the production of high quality electron bunches at 8 GeV from a single stage, the staging of two LPA modules, novel injection techniques for ultra-high beam brightness, investigation of processes that stabilize beam break up, new concepts for positron acceleration, and new technologies for high-average-power, high-efficiency lasers. In addition to the long term goal of a high energy collider, LPAs can provide compact sources of particles and photons for a wide variety of near-term applications in science, medicine, and industry.

Research on LPAs has exploded in recent years, driven in part by the extremely rapid advances made in high-power lasers based on the 2018 Nobel Prize winning technique of chirped-pulse amplification. Numerous high-power laser facilities have sprung up worldwide, particularly in Europe and Asia. Consequently, about 800--1000 research papers are published annually on LPAs. Since much of this research is overseas, it is critical that the U.S.\ make strong investments in LPAs to ensure global leadership.

The LPA community proposes the following recommendations to the Snowmass conveners:

\begin{enumerate}
\item Vigorous research on LPAs, including experimental, theoretical, and computational components, continue as part of the General Accelerator R\&D program to make rapid progress along the LPA R{\&}D roadmap towards an eventual high energy collider,  develop intermediate applications, and ensure international competitiveness.

\item Enhance {R\&}D on laser drivers to develop the efficient, high repetition rate, high average power laser technology that will power LPA colliders.

\item 
Near-term LPA capability extensions should be carried out, such as enhancing existing facilities in laser performance, and the construction of a new facility for high repetition rate (kHz) lasers to enable precision LPA R{\&}D towards collider performance requirements.

\item An integrated design study of a high energy (1--15 TeV) LPA-based collider be performed that details all the components of the system, such the injector, drive laser, plasma source, beam cooling, and beam delivery system. This would be followed by a conceptual design report.

\item A study be carried out for a collider demonstration facility at an intermediate energy (20--100~GeV) that would demonstrate essential technology and provide a facility for physics experiments at intermediate energy.   

\item A DOE-HEP workshop be held in the near term to update and formalize the U.S.\ advanced accelerator strategy and roadmaps.

\end{enumerate}


\newpage
\end{widetext}

\tableofcontents

\section{\label{sec:intro} Motivation and Overview}%

The quest to extend the reach of particle colliders to ever higher energies is constrained by the enormous size and cost of these future machines. Consequently, there is a strong need to explore advanced accelerator technologies that can provide ultra-high accelerating fields, thereby reducing the size and cost of future colliders. 

Laser-driven plasma-based accelerators \cite{Esarey09,Hooker13}, also referred to as laser-plasma accelerators (LPAs),  can sustain accelerating fields of 10-100 GV/m, some three orders of magnitude beyond conventional RF linacs. These are the highest accelerating fields produced by any of the widely researched advanced accelerator concepts. This would enable LPAs to produce multi-TeV leptons in a relatively short distance, on the order of 1 km.

In addition, LPAs intrinsically produce very short particle bunches, owing to the correspondingly short wavelength (10-100 $\mu$m) of the plasma wave that accelerates the particles, which is some three orders of magnitude shorter than that of RF linacs. A short bunch length can significantly reduce beamstrahlung during the bunch collision, which ultimately leads to a reduction of overall power consumption by allowing operation at higher charge per bunch. 

The compact size of LPAs has implications towards the reuse of a future facility infrastructure. A future linear collider facility based on RF technology (e.g., a Higgs factory) could be re-purposed with LPA technology to provide higher particle energy, thereby saving on construction and infrastructure costs. 

Novel energy recovering methods are also possible with LPAs. Additional laser pulses can be used to absorb the residual energy in the plasma that is not transferred to the lepton beam. Owing to the high efficiency of photo-voltaic cells, unused laser energy can be recycled. Furthermore, plasma-based beam dumps can be used to recover much of the lepton beam energy after the collision point. These methods lead to reductions in the overall collider power consumption.

Research on LPAs is very active worldwide owing to the extremely rapid evolution of the laser technology driving LPAs.  LPA drive lasers are based on Chirped Pulse Amplification (CPA), the invention of which won the Nobel Prize for Physics in 2018 \cite{nobelprize18}. Presently, most high power CPA systems are based on Ti:Sapphire technology that is capable of producing multi-PW peak powers at a few Hz repetition rates with relatively low efficiencies. These systems have proved useful for laboratory demonstrations of the key LPA physics, however, an LPA-based collider requires much higher repetition rates and efficiencies. 
To meet these needs, new laser technologies are being developed, such as those based on coherent combination of fibers or Thulium, which in principal are capable of providing the high rep-rates (tens of kHz) and high efficiencies (tens of percent) needed for an LPA collider, as described in Sec.~\ref{sec:lasers}.

The flagship DOE research facility in the U.S.\ performing R\&D on LPAs is the BELLA Center at LBNL \cite{bella}. The centerpiece of this facility is the BELLA PW laser, which delivers 1~PW (40~J in 40~fs) pulses at 1~Hz. The main focus of the BELLA Center research program is development of single-stage  multi-GeV LPA accelerator modules, and the efficient coupling (or staging) of LPA modules at the multi-GeV level. The path to extending the particle energy is in staging, i.e., coupling together, many multi-GeV plasma modules, each driven by a separate synchronized laser driver, to realize a very high energy linac. The current world record for the electron beam energy from a single LPA stage is 8 GeV using the BELLA PW laser. To address the need for higher rep-rates, the kBELLA (kHz BErkeley Lab Laser Accelerator) facility has been proposed, which would house a 1 kHz, J-class laser system for precision LPA research, and would provide access for a wide variety of user-based science.  

The vast majority of LPA research worldwide has focused on the physics of the LPA electron linac, i.e., the physics of electron acceleration within the laser-driven plasma wave. Moving forward, to meet the needs of an integrated LPA collider study, much more emphasis needs to be placed on the other components of a future LPA collider, such as the positron linac, beam cooling systems, and beam delivery systems. For example, there is a need to develop compact cooling and delivery systems that are compatible with the intrinsically short particle bunches produced by LPAs. Some of the research needed for laying the foundation for an LPA collider is described in the 2016 Advanced Accelerator Roadmap \cite{roadmap16}. The advanced accelerator community strongly recognizes the need for an integrated LPA collider study and strongly encourages funding to support this effort. 

In addition to the long-term application of an LPA collider, LPAs provide an extremely compact accelerator technology for a wide range of near-term applications. These include LPA-driven light sources in the x-ray and gamma-ray regimes, as well as high brightness sources of electrons, protons, and ions. These intrinsically synchronized sources of photons and particles can enable a wide variety of pump-probe studies in the fields of basic and applied sciences. Near-term applications range from compact  accelerators for biology and medicine to probe beams for security and industry \cite{BRN19}.

The U.S.\ faces serious competition on laser and plasma accelerator research in Europe and Asia. Examples include the Extreme Light Infrastructure program \cite{eli} in Europe, which has constructed three new high-power laser labs with a budget in excess of one billion euro, the EUPRAXIA project \cite{Eupraxia-CDR} in Europe to explore precision plasma accelerators funded at the several million euro level, and several projects in China, Japan, and Korea on LPAs. For example, experiments at the Shanghai Institute of Optics and Fine Mechanics, China, have recently demonstrated an LPA-driven free-electron laser \cite{Wang21}.  In order for the U.S.\ to retain its leadership and remain competitive in these areas, a suitable level of funding is required. 

\begin{figure}
    \centering
    \includegraphics[width=0.9\columnwidth]{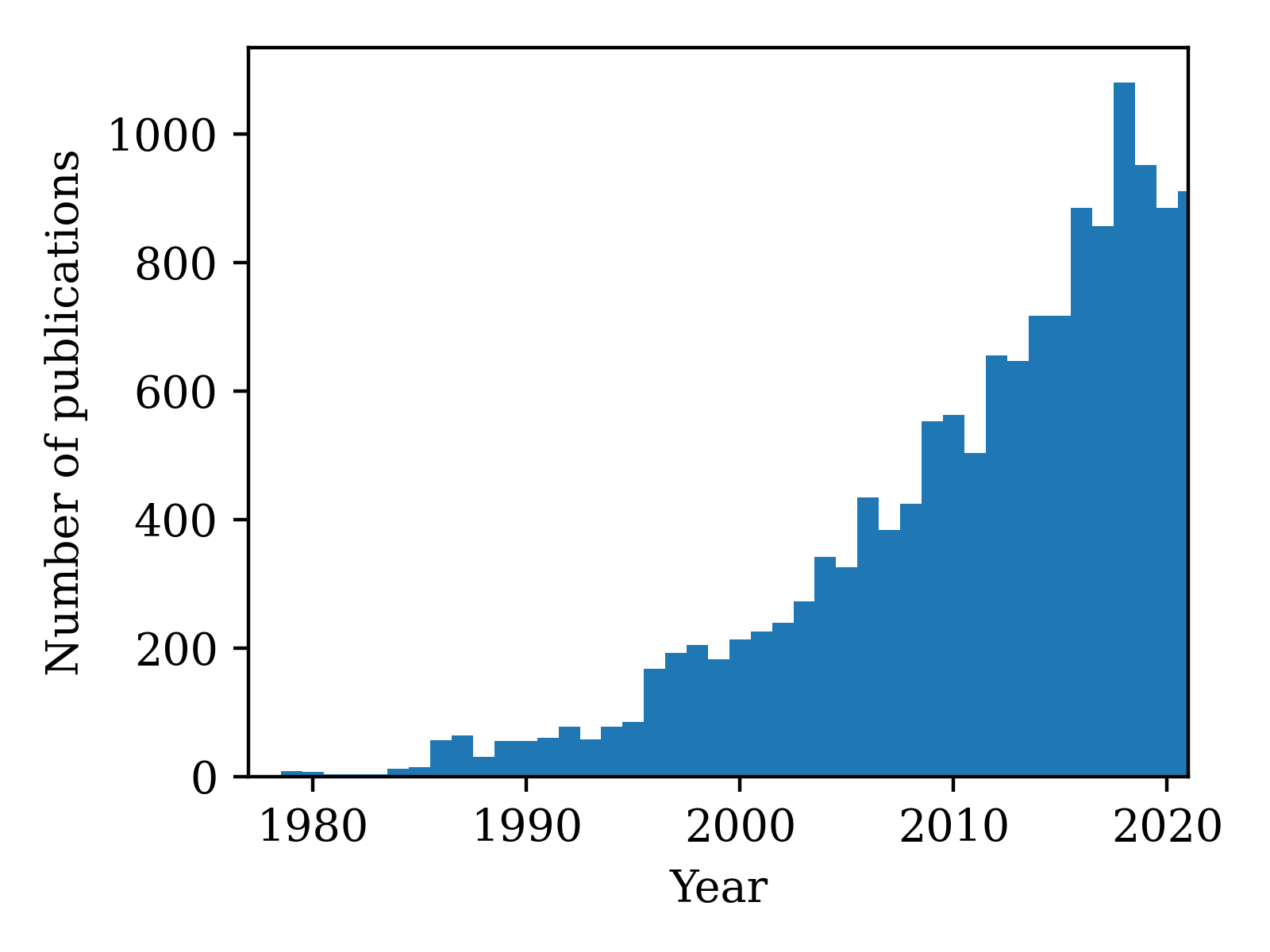}
    \caption{Number of publications per year as obtained from a Google Scholar search on articles containing all of the following keywords: ``laser$+$plasma$+$wakefield$+$accel*''.}
    \label{fig:nrofpubl}
\end{figure}

The widespread availability of high power laser facilities worldwide has led to an explosion of research results on LPAs in recent years. The number of publications per year on LPAs is shown in Fig.~\ref{fig:nrofpubl}, as compiled from a search in Google Scholar using the terms ``laser$+$plasma$+$wakefield$+$accel*''. Research results on LPAs are being obtained at an astounding rate with 800--1000 publications per year. This is further evidence that the state-of-the-art on LPAs is evolving at an extremely rapid pace, with new results emerging on a daily basis. This high level of activity on LPA research is certain to spawn numerous new ideas, concepts, techniques that will help bring a future LPA-based collider to fruition. 

\section{\label{sec:reports} Previous Reports and Recommendations}

Strategy updates developed in both Europe and the U.S. in the past decade prioritize R\&D on advanced accelerator concepts and laser-plasma accelerators.
For example, the 2014 {\it Report of the Particle Physics Project Prioritization Panel (P5)} \cite{P5-2014} highlights the importance of high-gradient acceleration technologies for future colliders (page~47):
\begin{quote}
    There is a critical need for technical breakthroughs that will yield more cost-effective accelerators. For example, ultra-high gradient accelerator techniques will require the development of power sources (RF, lasers, and electron beam drivers) compatible with high average power and high wall plug efficiency, and accelerating structures (plasmas, metallic, and dielectric) that can sustain high average power, have high damage threshold, and can be cascaded. Engagement of the national laboratories, universities, and industry will be essential for comprehensive R\&D to meet these challenges. Advancing these technologies will benefit many other areas of science and technology.
\end{quote}
Additionally the report \cite{P5-2014} highlights plasma acceleration as a potential technology to  upgrade future electron-positron colliders (Recommendation~24 on page~20): 
\begin{quote}
A multi-TeV e+e– collider could be based on either the Compact Linear Collider (CLIC) or plasma-based wakefield technology. The wakefield technology would be done as an energy upgrade to the ILC, or located elsewhere.
\end{quote}
Following the 2014 P5 Report \cite{P5-2014}, the Accelerator Research and Development Subpanel was formed and made recommendations for the DOE General Accelerator R\&D (GARD) program in the 2015 report {\it Accelerating Discovery: A Strategic Plan for Accelerator R\&D in the U.S.} \cite{HEPAP15}.  This report stressed the need to adequately support laser-driven plasma wakefield acceleration R\&D towards an HEP collider as well as required demonstration facilities (pages 11 and 30). 

Similarly, the need to intensify development of high-gradient and plasma-based accelerator technology was recommended in the recent report {\it 2020 Update of the European Strategy for Particle Physics} \cite{ESPPU20} (page~9):
\begin{quote}
    Innovative accelerator technology underpins the physics reach of high-energy and high-intensity colliders. It is also a powerful driver for many accelerator-based fields of science and industry. The technologies under consideration include high-field magnets, high-temperature superconductors, plasma wakefield acceleration and other high-gradient accelerating structures, bright muon beams, energy recovery linacs. The European particle physics community must intensify accelerator R\&D and sustain it with adequate resources. A roadmap should prioritize the technology, taking into account synergies with international partners and other communities such as photon and neutron sources, fusion energy and industry. Deliverables for this decade should be defined in a timely fashion and coordinated among CERN and national laboratories and institutes.
\end{quote}
This 2020 European Strategy report \cite{ESPPU20} was followed by a report of the European {\it Laboratory Directors Group} \cite{LDG22} that developed an accelerator R\&D plan and concluded (page 137): 
\begin{quote}
The field of high-gradient plasma and laser accelerators offers a prospect of facilities with significantly reduced size that may be an alternative path to TeV scale e+e- colliders. Though presently at an earlier development stage than the other fields, first facilities in photon and material science are now feasible and are in preparation. These accelerators also offer the prospect of near term, compact and cost-effective particle physics experiments that provide new physics possibilities supporting precision studies and the search for new particles. 
\\
The expert panel has defined a long term R\&D roadmap towards a compact collider with attractive intermediate experiments and studies. It is expected that a plasma-based collider can only become available for particle physics experiments beyond 2050, given the required feasibility and R\&D work described in this report. It is therefore an option for a compact collider facility beyond the timeline of an eventual FCC-hh facility. A delivery plan for the required R\&D has been developed and includes work packages, deliverables, a minimal plan, connections to ongoing projects and an aspirational plan. The panel recommends strongly that the particle physics community supports this work with increased resources in order to develop the long term future and sustainability of this field.
\end{quote}

Development of laser-driven plasma accelerator technology is a high priority for the international HEP community. In addition to the HEP community, the following reports stressed their strong interest in and support of laser-driven plasma accelerator technology, highlighting synergies and possible technology spin-offs:
\begin{itemize}
    \item The 2019 report of the {\it Basic Research Needs Workshop on Compact Accelerators for Security and Medicine} \cite{BRN19} states strong interest in and support for compact 100 MeV to 100 GeV accelerators based on laser-driven wakefields (pages 44, 77, 78, 158, 193, 196, 201 and 202).
    \item The 2019 {\it Brightest Light Initiative Workshop} \cite{BrightestLightReprot18} report states that (page 3.18):
    \begin{quote}
        Extreme acceleration gradients in laser-plasma accelerators can be leveraged for future applications and light sources that need low-emittance, high-brilliance beams by investing in short-pulse laser systems with kHz to MHz repetition rates.
    \end{quote}
    \item The 2020 {\it Report of the Fusion Energy Sciences Advisory Committee} \cite{DPP20}, titled `Powering the Future Fusion \& Plasmas' (pages 18 and 35) recommends to (page 35):
    \begin{quote}
        Pursue the development of a multi-petawatt laser facility – and a high-repetition-rate high-intensity laser facility in the US, in partnership with other federal agencies where possible.
    \end{quote}
    \item The 2021 {\it Plasma Science, Decadal Report} \cite{NAS-Decadal20} put forward five recommendations encouraging collaborative research on plasma acceleration in theory, computation, and experiments (pages 158--160).

\end{itemize}

\section{\label{sec:status} Background and Current Status}

In an LPA, an intense laser pulse propagating in an underdense plasma ponderomotively drives an electron plasma wave (or wakefield) \cite{Esarey09,Hooker13}. The plasma wave has a relativistic phase velocity and can support large fields suitable for charged particle acceleration and focusing.  In the standard LPA configuration, often referred to as a laser wakefield accelerator (LWFA), the plasma wave is driven by a single, resonant, short-pulse laser.
A relativistic charged particle (electron or positron) beam  co-propagating behind the laser at an appropriate phase in the laser-excited plasma wave may be accelerated to high energy. Optimal wake excitation requires a pulse duration $\tau_L \sim \omega_p^{-1}$, and a relativistic intensity $I_L$ such that $a^2 \simeq 0.73 I_L[10^{18}\,{\rm W/cm}^2] (\lambda[\mu{\rm m}])^2 \sim 1$, where $\lambda$ is the laser wavelength. Here $\omega_p=(4 \pi n e^2/m)^{1/2} = 2\pi c/\lambda_p$ is the plasma frequency, where $n$ is the ambient plasma density ($m$ and $e$ are the electron mass and charge, respectively). In this case, the characteristic amplitude of the wakefields is on the order of $E_0[\hbox{V/m}]\simeq 96 (n[\hbox{cm}^{-3}])^{1/2}$. For example, for $n=10^{17}$ cm$^{-3}$ the accelerating gradient is $E_0\simeq 30$ GV/m, several orders of magnitude larger than in conventional RF accelerators. 

The energy gain in a singe LPA can be limited by several factors: laser driver diffraction, beam-wave dephasing, or laser driver depletion \cite{Esarey09}. Laser diffraction leads to a reduction of laser intensity during propagation of the laser and, hence, reduction of the accelerating gradient. Optical guiding by means of plasma channels can be used to mitigate driver diffraction and extend the laser-plasma interaction region beyond the characteristic laser diffraction length. However, since the wakefield phase velocity is on the order of the laser group velocity, a highly-relativistic particle beam will eventually outrun the accelerating and focusing region of the wake. The propagation length over which this occurs is the dephasing length.
Longitudinal plasma tapering may be used to extend the dephasing limit.  Alternatively, laser spatio-temporal couplings, e.g., a flying focus, has been proposed to control the phase velocity and extend the interaction beyond the dephasing length \cite{Palastro20}. Ultimately, the single LPA stage energy gain is determined by laser energy depletion--the laser deposits its energy into excitation of the plasma wave. The depletion length scales with the plasma density as $L_d\propto n^{-3/2}$ and, hence, the single stage energy gain scales as $E_0 L_d \propto n^{-1}$. Reaching high energies in a single LPA stage requires operating at low plasma densities. However, this results in a reduction of the average accelerating gradient, and requires increasing the laser driver energy so that the laser dimensions are longitudinally and transversely resonant with the plasma wavelength, $U_L\propto \lambda_p^3 \propto n^{-3/2}$.  
Increasing the beam energy can be achieved by cascading several LPA stages, each one powered by an independent laser pulse (staging). Staging of LPAs provides a path to high energy while keeping a high average accelerating gradient (assuming the inter-stage distance required to in-couple the laser and beam is kept small) and without increasing the laser pulse energy. 
An LPA-based collider would consist of multiple LPA accelerating stages \cite{Leemans09,Schroeder10b}.

Beams accelerated in an LPA are intrinsically short \cite{Buck11,Lundh11} compared to the ones used in conventional accelerators, typical bunch lengths are a fraction of the plasma wavelength, namely $L_b \lesssim \lambda_p$ (e.g., tens of micron for $n=10^{17}$ cm$^{-3}$). By properly shaping the beam current profile, the longitudinal dependence of the beam-loaded accelerating field can be controlled, minimizing energy spread growth during the acceleration.

Since the first demonstration of high-quality beams from LPAs in 2004 \cite{Geddes04,Faure04,Mangles04}, steady progress has been made increasing the beam energy gain in a single LPA stage. The current record LPA energy gain of $7.8$ GeV was obtained in 2018 by guiding a $0.85$~PW laser in a 20 cm-long (corresponding to about $15$ diffraction lengths) laser-heated capillary discharge waveguide \cite{Gonsalves19}. Development of laser guiding techniques that allow an LPA to operate at low plasma densities, while still keeping the laser driver tightly focused over distances much longer than its characteristic diffraction length are critical to the realization of high-energy LPAs. In this respect, several schemes for the production of meter-scale plasma waveguides using optical-field-ionization techniques have been recently proposed and validated \cite{Shalloo19, Miao20}.  It has also been shown that these techniques are capable of high repetition rate operation \cite{Alejo22}.

Staging of two independently-powered LPAs was first demonstrated in 2016 \cite{Steinke16}. In this proof-of-principle experiment, stable $\sim 100$ MeV electron beams from the first LPA (length $\lesssim 1$ mm) were focused by a discharge capillary-based active plasma lens of length 1.5 cm into a 3.3 cm-long LPA powered by a second laser pulse that was in-coupled into the second stage by means of a plasma mirror. The second LPA stage operated in a dark-current-free, quasi-linear regime. Acceleration by the wakefield of the second stage was observed via a $\sim 100$ MeV energy gain for a subset of the electron beam. Experiments aiming at demonstrating staging at the GeV and multi-GeV level with high capture efficiency ($> 90\%$ particles transported and accelerated in the second stage) and beam quality (emittance and energy spread) preservation are planned at the recently commissioned second beamline of the BELLA Center. Beam quality preservation during staging requires, among other things, development of thin-film plasma mirrors \cite{Zingale21}.

The required properties of future colliders will be determined by high-energy physics experiments that are currently underway.  However, it is anticipated that a center-of-mass energy $\gtrsim 1$ TeV and a luminosity $\gtrsim 10^{34}$  cm$^{-2}$s$^{-1}$ will be required. To minimize the power requirements, this implies using beams with high charge per bunch and low normalized  emittances $< 0.1$ $\mu$m.  In addition small relative energy spreads, $< 1\%$ are required in order to effectively transport the bunch between subsequent plasma stages without emittance degradation and to guarantee a sufficiently small bunch size at the interaction point. So far LPAs have demonstrated the production of high-quality electron beams with HEP-relevant parameters, such as relative energy spreads as low as $\sim$1\% \cite{Kirchen21,Rechatin09}, normalized emittances of $\sim$ 0.1~$\mu$m \cite{Plateau12,Weingartner12}, and high charge (100s of pC) \cite{cuperus17,Goetzfried20}, even though the best parameters were not all achieved, in general, simultaneously. Several controlled injection techniques (e.g., colliding-pulse injection \cite{Esarey97, Faure07, Rechatin09}, ionization-induced injection with one \cite{Chen10, Pak10, McGuffey10} or multiple pulses \cite{Yu14, Tomassini19}, and density gradient injection \cite{Bulanov98, Geddes08, Gonsalves11,Goetzfried20}), aimed at increasing the beam quality, stability, and tunability, have been proposed and, in some cases, experimentally implemented and their advantages demonstrated.

\section{\label{sec:LPA-LC} Laser-Plasma Linear Collider Concept}

A conceptual design of LPA-based linear collider has not been completed. However, several preliminary studies have been performed \cite{Schroeder10b,Schroeder16}. These studies have been used to help identify the expected operational plasma and laser parameters to guide the R\&D toward developing plasma-based collider concepts \cite{ANAR17}.  Many of the parameters are determined by the plasma density choice. The plasma density regime considered in these studies, operating at a plasma density of the order of $n\sim 10^{17}$~cm$^{-3}$, provides for a large accelerating gradient ($\propto n^{1/2}$) and reduces the linac power requirements ($\propto n^{1/2}$) , while keeping the beamstrahlung at the collision point at an acceptable level ($\propto n^{-1/2}$)  for a target luminosity.

\begin{figure*}
    \centering
    \includegraphics[width=1.4\columnwidth]{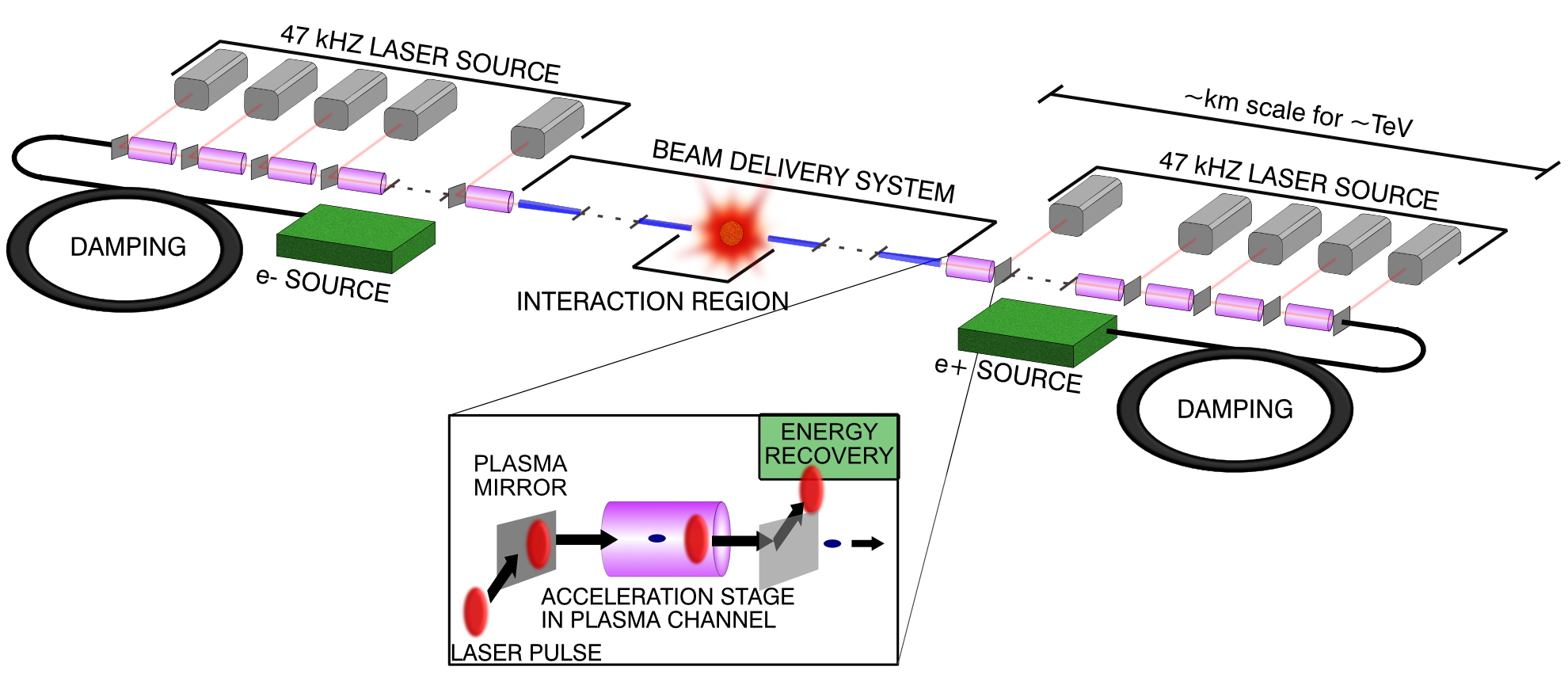}
    \caption{Schematic of an LPA-based linear collider.}
    \label{fig:cartoon}
\end{figure*}

It is envisioned that an LPA-based collider would initially operate at $\sim$TeV center-of-mass energy, and be upgraded to $>10$~TeV.  
Section~\ref{sec:example} presents potential $e^{+}e^{-}$ linear collider parameters for center-of-mass energies of 1 TeV, 3 TeV, and 15 TeV.  
A schematic of a LPA-based collider is shown in Fig.~\ref{fig:cartoon}.
In these examples, the same single-stage laser system and LPA stage is employed, enabling upgrading of the collider energy using existing collider infrastructure.  Each stage would be powered by a 50~TW peak power laser system, with 6.5~J per pulse, operating at 47~kHz repetition rate.  Table~\ref{tab:stage} lists the laser-plasma parameters of the LPA stage.  The 300~kW average laser power is beyond current technology; however, a development path has been identified to achieve these laser parameters, as discussed in Sec.~\ref{sec:lasers}.
These numerical examples are representative of what could be achieved using LPA-based linacs.  LPA technology is rapidly advancing, with new innovations appearing at a swift pace, and we fully expect that the LPA collider design and numerical parameters will evolve as new innovations and inventions in this field emerge.

\begin{table}
\caption{\label{tab:stage} LPA stage laser and plasma parameters }
\begin{ruledtabular}
\begin{tabular}{lc}
Laser pulse energy  & 6.5~J  \\
Laser (FWHM) pulse duration  & 130~fs \\
Laser pulse peak power & 50~TW\\
Laser wavelength & 1~$\mu$m \\
Plasma density  &  $10^{17}$~cm$^{-3}$\\
Plasma channel length & 1.7~m\\
Plasma channel radius & 22~$\mu$m \\
Peak accelerating field & 6~GV/m\\
Beam peak current & 3~kA \\
RMS beam length & 8.5~$\mu$m\\
Loaded accelerating gradient & 3~GV/m \\
Particle energy gain per stage & 5~GeV
\end{tabular}
\end{ruledtabular}
\end{table}

One may also consider a gamma-gamma collider using LPA electron beams.  In a $\gamma\gamma$ collider, two electron beam linacs would be employed and Compton backscattering used near the interaction point to generate colliding energetic photon beams.  Photon collisions can access many of the lepton interactions available in an $e^{+}e^{-}$ collider, and a $\gamma\gamma$ collider eliminates the need for positron beam generation and acceleration in plasma, and removes the beamstrahlung and beam-beam constraints.  Removing beamstrahlung constraints, allows a larger charge per bunch and power savings to reach a given luminosity.  In addition, round beams may be considered. Since only electrons are required, one may operate in the nonlinear (bubble) regime of laser-plasma acceleration, which has some advantages over the quasi-linear regime.  Section~\ref{sec:gg-collider} presents parameters for a  $\gamma\gamma$ collider using LPA electron beams in the nonlinear regime.

\subsection{Example: electron-positron collider with $\sqrt{s}= $1~TeV, 3~TeV, and 15~TeV}
\label{sec:example}

In this section, the numerical examples of collider parameters are based on LPAs operating in the quasi-linear regime using plasma channels \cite{Schroeder16}.  Plasma channels are employed for guiding the intense laser pulses.   
Operating at a density of $n=  10^{17}$~cm$^{-3}$ and with a channel radius of $22~\mu$m, a 6.5~J, 130~fs FWHM duration laser pulse can excite a plasma wave with peak accelerating field of 6~GV/m. An appropriately shaped (triangular current distribution) and phased particle beam with 0.19~nC of charge (3~kA beam peak current) can be accelerated by 5~GeV over 1.7~m, with a loaded gradient of 3~GV/m.  This stage has 75\% wake-to-beam efficiency and 20\% laser-to-wake efficiency.  Energy recovery options for the laser energy as well as the energy remaining in the plasma wave may be employed.  In these numerical examples partial recovery of the remaining laser energy (after the laser-plasma interaction) using photo-voltaics is assumed to improve the overall energy efficiency. 

For this illustrative numerical example, hollow plasma channels were considered \cite{Schroeder16}.  Hollow plasma channels enable symmetric acceleration of electron and positron beams, at beam densities larger than the ambient electron plasma density.  Very similar performance as that presented in Table~\ref{tab:stage} may be achieved for electron acceleration using other plasma channel geometries operating at that density.   Although here we are considering symmetric electron and positron beam linacs (using the same laser systems and plasma channel technology), the electron and positron linacs could be optimized at different operating points, possibly using different laser and plasma channel technologies. 

Table~\ref{tab:collider} shows the high-level collider parameters.  A number of assumptions were made in Table~\ref{tab:collider}. It assumed flat beams (asymmetric vertical and horizontal emittances), with vertical focusing at the IP of $\beta^* = 0.1$~mm to yield $\sim$nm-scale beam sizes at the IP.  The details of the beam delivery system and final focus configuration are to be determined, as well as the beam cooling system to achieve the emittances required for the IP spot sizes.  Note that initial studies indicate that beam depolarization during the acceleration in plasma accelerators is low for collider-relevant beam emittances and fulfills the requirements for high energy physics experiments \cite{Vieira11}.

In Table~\ref{tab:collider} the stated linac length is for each arm of the accelerator.  The AC power listed in Table~\ref{tab:collider} is for acceleration in both of the two linac arms.  The overall wall-to-laser efficiency was assumed to be 50\%.  This laser efficiency is  challenging, but recent R\&D (see Sec.~\ref{sec:lasers}) indicates that this is technically possible by coherent combining of fiber lasers with electrical-to-optical efficiency of the diode-pump lasers $\sim$65\%, the optical-to-optical efficiency of the fiber lasers $\sim$90\% (owing to the low
quantum defect), and the efficiency of combining/stacking fibers $\sim$85\%. 

\begin{table}
\caption{\label{tab:collider} High-level electron-positron collider parameters}
\begin{ruledtabular}
\begin{tabular}{lccc}
Center-of-mass energy [TeV] & 1 & 3 & 15 \\
Beam energy [TeV] & 0.5 & 1.5 & 7.5\\
Luminosity [$10^{34}$ cm$^{-2}$ s$^{-1}$] & 1 & 10 & 50\\
Particles/bunch [$10^9$] & 1.2 & 1.2 & 1.2\\
Beam power [MW] & 4.4 & 13 & 65 \\
RMS bunch length [$\mu$m] & 8.5 & 8.5 & 8.5 \\
Repetition rate [kHz] & 47 & 47 & 47\\
Time between collisions [$\mu$s] & 21 & 21 & 21 \\ 
Beam size at IP, x/y [nm] & 50/1 & 10/0.5 & 4/0.25\\
Linac length [km] & 0.22 & 0.65 & 3.3 \\
Facility site power (2 linacs) [MW] & 105 & 315 & 1100 \\
\end{tabular}
\end{ruledtabular}
\end{table}

\subsection{Example: gamma-gamma collider with $\sqrt{s}=15~{\rm TeV}$}
\label{sec:gg-collider}

In this section we present an example of a $\gamma\gamma$ collider using electron beams accelerated by LPAs in the nonlinear regime.  There are several regimes of laser-driven plasma acceleration that may be accessed based on the intensity of the laser pulse.  Section~\ref{sec:example} presents collider designs based on operation in the quasi-linear regime.    For high laser intensities, the LPA can operate in the bubble regime, where (almost) all the electrons are expelled by the laser ponderomotive force, forming an ion cavity co-propagating behind the laser.  In the bubble regime, the accelerating field is independent of the transverse position and the focusing field is linear with respect to the transverse coordinate and independent of the axial position (conserving the electron beam transverse normalized rms emittance).  
Note that the transverse fields in the ion cavity are defocusing for positrons; hence, stable positron acceleration is problematic in the nonlinear regime in a uniform plasma.  Wakefield excitation in plasma columns have been proposed for modifying the wakefield to allow for positron focusing and acceleration \cite{Diederichs20}.   In the bubble regime, the laser effectively creates a plasma channel and can self-guide over a distance corresponding to many Rayleigh ranges.  

Table~\ref{tab:bubble-stage} shows an example of single-stage LPA parameters operating in the bubble regime.  This single-stage LPA example is based on PIC modeling of the nonlinear laser-plasma interaction \cite{Schroeder22}.  The laser energy depletion at the end of the stage is 20\%. (In principle, the majority of the remaining laser energy could be recovered with a photo-voltaic.) The wake to beam energy efficiency of this example is 43\%.

\begin{table}
\caption{\label{tab:bubble-stage} LPA stage laser and plasma parameters, operating in the nonlinear bubble regime}
\begin{ruledtabular}
\begin{tabular}{lc}
Laser pulse energy  & 50~J  \\
Laser (FWHM intensity) pulse duration  & 70~fs \\
Laser spot size  & 31~$\mu$m \\
Laser strength parameter, $a_0$ & 4.5 \\
Laser pulse peak power & 0.43~PW\\
Laser wavelength & 0.8~$\mu$m \\
Plasma density  &  $4.6 \times 10^{17}$~cm$^{-3}$\\
Plasma cell length & 3.1~cm\\
Bunch charge & 1.2~nC \\
Bunch number & $7.5\times 10^9$\\
RMS beam length & 2.2~$\mu$m\\
Loaded accelerating gradient & 117~GV/m \\
Particle energy gain per stage & 3.2~GeV
\end{tabular}
\end{ruledtabular}
\end{table}

Table~\ref{tab:gg-collider} presents the collider parameters for a $\gamma\gamma$ collider based on the LPA stage parameters in Table~\ref{tab:bubble-stage}.   Here we consider round beams, with 100~nm normalized transverse emittance in each plane.  Such electron beams could be generated via all-optical plasma-based injection techniques, as described in Sec.~\ref{sec:subsys}, removing the need for damping rings.  
The basic approach to a $\gamma\gamma$ collider is described in Ref.~\cite{Telnov95}.
Maximizing the photon energy while avoiding pair creation during scattering requires 
the incident photon wavelength to satisfy $\lambda [\mu\textrm{m}] \simeq 4 U_b$[TeV],  where $U_b$ is the beam energy, and the peak scattered photon energy is $\hbar \omega \simeq 0.82 U_b$[TeV]. To achieve 7.5~TeV photons requires a 9.1~TeV electron beam scattering off 0.0345~eV photons. With a conversion efficiency (scattered photons per electron) of $N_\gamma/N_e = 0.63$, approximately 0.7~TW of power is required in the scattering radiation pulse.  The geometric luminosity of the photon beams is reduced compared to the electron beams,  $\mathcal{L}_\gamma = (N_\gamma/N_e)^2 \mathcal{L}_e$.
Development of high-average-power radiation sources for scattering in the mid-IR regime presents a technical challenge to development of $\gamma\gamma$ colliders operating at multi-TeV energies.   As Table~\ref{tab:gg-collider} indicates, the high charge per bunch and round beams greatly reduces the collider power requirements.

\begin{table}
\caption{\label{tab:gg-collider} High-level $\gamma\gamma$ collider parameters}
\begin{ruledtabular}
\begin{tabular}{lc}
Center-of-mass photon energy  & 15~TeV \\
Photon beam energy & 7.5~TeV\\
Electron beam energy  & 9.1~TeV\\
Beam power & 12.5~MW\\
Repetition rate  & 1.1~kHz \\
Time between collisions  &  0.9~ms\\ 
Vertical beam emittance & 100~nm\\
Horizontal beam emittance & 100~nm\\
Final focus beta functions, $\beta_x^*=\beta_y^*$ & 0.15~mm \\
Scattering wavelength & 36~$\mu$m\\
Geometric luminosity  & $2\times 10^{35}$~cm$^{-2}$ s$^{-1}$\\
Facility site power (2 linacs)  & 415~MW\\
\end{tabular}
\end{ruledtabular}
\end{table}

\subsection{Upgrade of a future RF collider facility}

Owing to the ultra-high geometric gradients an LPA-based linac would be sub-km to reach several TeV beam energies.  For example, a geometric gradient of 2~GV/m, requires potentially only 0.5~km in each linac arm to reach 1~TeV beam energies.  This offers the possibility of re-using the tunnels of an existing conventional linear collider complex to greatly extend the energy reach of the facility.   
For example, one could consider using the ILC infrastructure \cite{ILC-TDR} and upgrading the beam energy by replacing the superconducting RF cavities with LPA stages.  The laser drivers occupy an area $\sim$m$^2$ and can fit in the ILC main linac tunnel.
Upgrading to $\sqrt{s}=$1~TeV or 3~TeV, with the parameters in Table~\ref{tab:collider} could be supported by the ILC site power.  The LPA beam power for $\sqrt{s}=$1~TeV and 3~TeV would be 4~MW and 12~MW, respectively. These values are within the power rating of the planned ILC beam dump.  The unused main linac tunnel length could be employed to extend the BDS system to accommodate the higher beam energies, as well as space for linear cooling sections to further reduce the beam emittance.  The bunch structure employed by the LPA linacs is one bunch each $\sim$20~$\mu$s, and additional bunch compressors would be required to achieve the short, 10-$\mu$m-scale, bunch length.  Furthermore, achieving higher beam energies is straightforward by adding additional LPA stages, although the required increased luminosity would require site power beyond the planned ILC design.  This provides a long-term upgrade path to continue realizing new physics reach in realistic stages using the infrastructure of a RF linear collider facility.  

\section{Research and development path toward a TeV collider}

Significant research is required to develop both the laser-plasma accelerator technology and the auxiliary systems compatible with laser-plasma-accelerated particle beams.  In addition, laser technology must advance to achieve the required high peak and average power short pulse lasers.  In this section, we discuss the required R\&D to construct a multi-TeV LPA-based collider.  An R\&D roadmap toward an LPA-based collider was described in the  2016 U.S.\ DOE \emph{Advanced Accelerator Development Strategy Report} \cite{roadmap16}.  In Sec.~\ref{sec:roadmap} we describe a proposed updated version of the R\&D roadmap.

In the next ten years, LPA research should focus on demonstration of staging of multi-GeV LPAs---the building block of a laser-plasma-based linac.  Also in the next decade, laser technology research should deliver kW average power Joule-class laser systems, providing a proof-of-principle demonstration of the path towards the 100~kW-class systems required for collider applications.  In parallel, an integrated design study should be completed to understand the required collider auxiliary systems (e.g., beam delivery system, positron source, injector, cooling system, etc.) compatible with the LPA linacs.

New experimental facilities will be required to carry out this research.  To address the need to develop LPA technology at high repetition rates, the kBELLA facility has been proposed, which would house a 1 kHz, J-class laser system for precision LPA research.  Such a facility would not only provide a proof-of-principle demonstration of the laser technology, but would provide a platform to develop high-repetition-rate LPA components (e.g., plasma sources, targetry, feedback systems, diagnostics, etc.).  

Beyond the next decade, it is envisioned that a low-energy (few tens of GeV) collider demonstrator would be required.  Given the cost of such a facility, a robust science case  needs to be developed (e.g., such an energy range may have potential applications to nuclear physics and QCD experiments).  With sufficient funding, such a facility could be constructed and commissioned on the 20-year timescale.   It is anticipated that, with sustained funding, a technical design report for a multi-TeV collider could be completed in the 25+ year timescale. 

As the laser-plasma-based accelerator technology develops, intermediate applications will become available that will have societal benefits as well as benefiting other DOE offices.   Owing to the compact nature of the laser drivers, laser-plasma accelerators also have the
potential to revolutionize accelerator-based science at lower energies, e.g., as drivers for compact free-electron lasers, advanced x-ray mono-energetic photon sources, and medical accelerators.  

\subsection{Laser-plasma accelerator R\&D }

The key building blocks of an LPA linac are the development of a robust multi-GeV LPA stage and the staging of multiple LPAs to reach high energy. Development of a robust multi-GeV LPA stage entails development of plasma channel creation techniques that enable laser guiding and plasma density tapering techniques to control and tune the plasma wave phase velocity.  Laser longitudinal and transverse shaping will be required to ensure mode quality and resonant excitation.  Particle beam current profile shaping will also be required to optimize beam loading, enabling a high wake-to-beam energy extraction efficiency. Minimizing the chromatic emittance growth associated with beam energy spread is a critical issue, and plasma-based beam phase space manipulation methods should be explored to address this issue. 

Multi-GeV staging is the highest near-term research priority for LPA development.  Proof-of-principle LPA staging has been performed at the 100-MeV energy level \cite{Steinke16}.  Critical to staging is the development of plasma mirror technology compatible with high-rep-rate operations that allows for compact laser coupling in and out of LPA stages.  Ultra-thin ($\sim$20~nm), replenishable, liquid crystal plasma mirrors \cite{Zingale21} are a promising candidate to provide compact laser coupling.  Also key to staging is development of compact methods to couple the particle beam between stages, minimizing beam loss and preserving beam emittance.  

Current high-peak-power laser systems operate at a few Hz repetition rates, and laser-plasma acceleration techniques, including plasma targetry, laser shaping, diagnostics, etc., will need to be developed for high repetition rates (tens of kHz). Laser-plasma acceleration of electron beams operating at 1~Hz have displayed significant shot-to-shot variations in electron beam parameters \cite{Maier20}. Air and ground motion at a frequency of order 100~Hz has been shown to  generate fluctuations that limit the precision of current laser systems.  At kHz and higher repetition rates, active feedback (utilizing machine learning and artificial intelligence methods) will enable unprecedented stability, control, and precision.   Developing a kHz LPA will be a key step toward application of this accelerator technology. 
  
Experimental demonstration of laser-plasma-based positron acceleration is required.  With a $\sim$10~GeV LPA stage, positron beams may be generated, via bremsstrahlung photo pair creation.  Such beams could be captured and injected into a staged LPA, enabling a compact method to investigate and perform experiments on LPA-based positron acceleration and transport. 

In addition to development of a multi-GeV LPA stage and coupling of stages, advanced laser-plasma injection techniques should be pursued to generate ultra-low emittance ($<$100~nm) electron beams.  Compact all-optical laser-plasma-based electron injectors may be considered for the collider injector (see Sec.~\ref{sec:subsys}).  Development of ultra-high-brightness electron sources would benefit a wide range of applications beyond high-energy physics.

\subsection{High-power laser R\&D }
\label{sec:lasers}

An LPA-based linear collider and other intermediate applications of plasma-based accelerator technology  will require laser drivers with repetition rates ranging from kHz to tens of kHz at Joules of pulse energies and high efficiency, a significant step from current state-of-the-art lasers. Additionally, a collider would require hundreds of such laser systems, one powering each accelerator stage. A detailed discussion of the development strategies of laser technologies that could achieve the required laser parameters is presented in a white paper submitted to Snowmass21 by the laser community \cite{SnowmassLaser21}.

In the near-term, diode-pumped, actively-cooled Ti:Sapphire systems \cite{Sistrunk17} offer a path to kHz rates. This will enable active correction \cite{Workshop13,Workshop17,falcone20} such as pointing, pulse energy and shape, and machine learning optimization \cite{He15,Galvin18} to realize the potential of precision LPAs. One approach under consideration uses an OPCPA front end, a cryogenically-cooled Ti:Sapphire, and incoherently combined fiber pump lasers, with the potential for further power scaling by increasing the pump rate to multi-kHz with thousands of fiber-based pump lasers \cite{Workshop17}.

In the mid- and long-term, to reach collider-relevant luminosities, laser repetition rates would need to increase to tens of kHz and efficiencies to tens of percent.  This requires new laser technologies because of the fundamental limits of the Ti:Sapphire properties. Multiple diode-pumped solid-state laser technologies have been proposed as mid- to long-term solutions towards higher-efficiency, higher-rate laser drivers in the sub-100~fs, few J regime, and all need sustained R\&D to push beyond current laser limitations and reach technical readiness for application as LPA collider drivers \cite{Workshop13,Workshop17,falcone20}. The fiber laser solution, coherently combining many ultra-short pulses generated from fiber amplifiers in time, space, and wavelength \cite{Zhou:15,Zhou:18,Chang:13}, offers high wall-plug efficiencies and excellent thermal management. A flexible high-rate laser driver utilizing bulk Tm:YLF crystals \cite{Galvin19,Siders19} near 2~$\mu$m wavelength is an inherently high-efficiency, single laser beam solution. 
Over the long term, the development of reliable, high-rep-rate sources in the mid-wave infrared (3--8$\mu$m) and long-wave infrared (8--15~$\mu$m) could also provide unique opportunities for HEP applications, including as an enabling technology for an ``all-optical'' source of high-brightness, potentially spin-polarized, electron beams using two-color injection \cite{Yu14,Schroeder14,Tomassini19}. 
Other promising paths include using diode-pumped, cryogenically-cooled, thin-disk Yb:YAG or Tm:YLF as amplifier gain media \cite{Baumgarten:16,Russbueldt:10,Klingebiel:15,Dominik:21}, combined with nonlinear pulse
compression technologies \cite{Viotti:22,Fan:21}, or novel, plasma-based methods for modulating ps-duration pulses \cite{Jakobsson21}. All candidate laser technologies require development of active control techniques for laser precision and stability \cite{Bagnoud:04,BAHK201445}, and high peak and average power handling optics, including compression gratings \cite{ALESSI2019239,Power:20} and robust optical coatings \cite{Mitchell:15,Rethfeld_2017,Wu14}.

\subsection{Positron sources}
\label{sec:e+}

Development of compact, LPA-based sources for positrons is a key step in the  R\&D effort towards development of a TeV collider. 
Conventionally, positrons have been produced via bremsstrahlung radiation decay or via the conversion of a gamma photon into an $e^-e^+$ pair when it travels in the field of an atomic nucleus \citep{chao2013handbook}.
Bremsstrahlung radiation can be obtained by colliding a relativistic LPA-generated electron beam with an energy $E_b \geq 10\, \GeV$ on a high-Z element solid (typically W or Ti).  LPA electron beams with energies  $\sim 150$~GeV could be used in a magnetic undulator to emit photons with average energies $E_\gamma \sim 10 \,\MeV$ that in a high-Z target decay into $e^-e^+$ pairs. Both methods are suitable, in principle,  to be implemented in ultra-compact, all-optical setups using LPA accelerated electron beams.
In addition, the interaction of high-intensity lasers ($I_L \gtrsim 10^{22}\,\textrm{W}/\textrm{cm}^2$) with electron beams with $E_b \sim 10\, \GeV$ is in a regime where positrons can also be created via multi-photon Breit-Wheeler interaction \citep{zhang2020relativistic}.   Positrons production by any of the above-mentioned methods will yield beams with relatively poor phase space properties (i.e., large energy spread and emittance) and will require beam cooling. The development of advanced, compact, positron generation and cooling techniques could potentially benefit any future electron-positron collider.

\subsection{Auxiliary systems R\&D}%
\label{sec:subsys}

A collider based on LPA-linacs will require auxiliary systems compatible with laser-plasma accelerators, and in particular the beam time structure: single bunch durations of tens of femtoseconds, separated by tens of microseconds. 
R\&D on auxiliary LPA-collider systems is in a nascent state.  A critical advance will be the development of an integrated design study for an LPA-based collider that will include preliminary designs for all the collider subsystems.  Key components include the electron source, positron source, cooling system, bunch compressors, and the beam delivery system (BDS) to the interaction point (IP).  

Compact positron sources are discussed in Sec.~\ref{sec:e+}.  A conventional electron source could be considered, with additional bunch compression and shaping to produce the 10-micron-scale bunch lengths required for efficient beam loading in plasma accelerators.  All-optical, plasma-based injectors may also be considered.  Compared to conventional injectors,
all-optical plasma-based injectors are able to generate ultra-low emittance electron beams directly into the accelerating and focusing phase of a plasma wave.  Among the several techniques available,
injection via a plasma density transition, or downramp injection \citep{Bulanov98, Geddes08, Gonsalves11, Swanson17, Goetzfried20}, and ionization injection \citep{Chen10, McGuffey10, Pak10} have shown the production of tunable and controllable beams in accessible experimental setups.  Bunches generated with such techniques can achieve low normalized emittances ($\sim$100~nm) and energy spreads on the percent level \citep{barber2018parametric,Weingartner12,Plateau12}.  By decoupling the wake driver and the laser used for ionization injection, ultra-low emittance ($\sim$10~nm) electron beams may be generated \cite{Hidding12,Yu14,Schroeder14}.  In the all-optical method of ultra-low emittance beam generation, two lasers with different colors are employed \cite{Yu14,Schroeder14} or multi-pulse trains for wakefield excitation \cite{Tomassini19}.  Production of spin-polarized electron beams is also possible via laser ionization injection of electrons from certain atoms \cite{Nie21}. 
Development of sources of ultra-high brightness electron beams will benefit many applications beyond the collider.  A white paper discussing plasma-based sources of particle beams has been submitted to Snowmass21 \cite{e-source-Snowmass21}.

Cooling techniques compatible with a plasma-based collider will need to be developed. Cooling is conventionally accomplished by using radiation generation by permanent magnetic undulators followed by acceleration. 
Another possibility for radiation generation is to use a laser-excited plasma wave as an undulator \cite{Rykovanov15,Wang17}. 
Plasma-based radiators and rapid acceleration by LPAs could potential enable compact cooling at higher beam energies, and development of these advanced techniques could potentially benefit any future electron-positron collider.

A BDS is necessary to transport particle beams from the linac to the interaction point.  
A BDS for an LPA collider must be compatible with the ultra-short bunch durations and the sub-percent beam energy spread ($<$1\%). 
Ultra-short beams suppress beamstrahlung in the quantum beamstrahlung regime; however, at multi-TeV center-of-mass energies, flat beams may also be employed for futher beamstrahlung reduction and power savings.  A significant challenge is developing a compact BDS for multi-TeV beams that is comparable in size to the plasma-based linacs.  The topic of advanced beam delivery systems for advanced accelerators is discussed in detail in a white paper submitted to Snowmass21 \cite{BDS-Snowmass21}.

\subsection{Simulation tools}

Development of enhanced modeling capabilities is required in order to guide the R\&D effort towards a TeV collider. The numerical modeling, in 3D, of LPAs using conventional Particle-In-Cell (PIC) codes is a computationally challenging task \cite{Vay21}. This is due to the great disparity among the length scales involved in the simulation, ranging from the micron scale of the laser wavelength (e.g., electron quivering in the laser field) to the meter scale of the laser-plasma interaction length for a multi-GeV-class LPA stage. Modeling a collider requires simulating a chain of hundreds of such LPA stages together with all the optics required for inter-stage beam transport and laser driver in-coupling. In addition, ensemble runs of simulations on large parameter space will be required to estimate tolerances and assess the impact of non-ideal effects (e.g., misalignment, tilts, etc.). Addressing these challenges requires developing an array of novel, high-fidelity and fast numerical tools, and exploiting high-performance computing on the upcoming exascale-capable supercomputers. Examples of such tools are, for instance, the quasi-static approximation \cite{Sprangle90, Mora97}, use of reduced dimensionality models (axisymmetric or azimuthal mode decomposition) \cite{Mora97, Benedetti10, Lehe16}, use of the Lorentz-boosted frame method \cite{Vay07}, use of averaged ponderomotive force \cite{Mora97, Benedetti17b, Benedetti10, Terzani19, Terzani21}, and mesh refinement \cite{WarpXEAAC2017,WarpXAAC2018}.  

In addition to kinetic modeling, development of fast machine learning tools should be addressed and such instruments used to guide large-scale parameter scans and to help control the laser plasma accelerator \cite{Vay21, Jalas21, Shalloo20}.
Furthermore, in order to perform an end-to-end modeling of the collider, it is desirable to integrate into the simulations physics models that go beyond the conventional kinetic laser-plasma physics description.  For example, incorporating plasma hydrodynamics for long-term plasma/gas evolution and accurate plasma/gas profiles description, plasma ionisation/recombination, models for heat transport in and out of the plasma source, coupling with conventional beamlines, spin tracking, production of secondary particles, radiation reaction and photon production (including single quanta events).

The development of simulation tools needed for the design of a multi-TeV collider will require robust and sustained team efforts based on collaborations in the accelerator modeling community, as well as coordination between national laboratories and university groups. These collaborations should leverage the past, ongoing, and future efforts from the DOE SciDAC and the Exascale Computing Project.

\subsection{Updated R\&D roadmap}
\label{sec:roadmap}

\begin{figure*}[htbp]
\centering 
\includegraphics[width=0.8\textwidth]{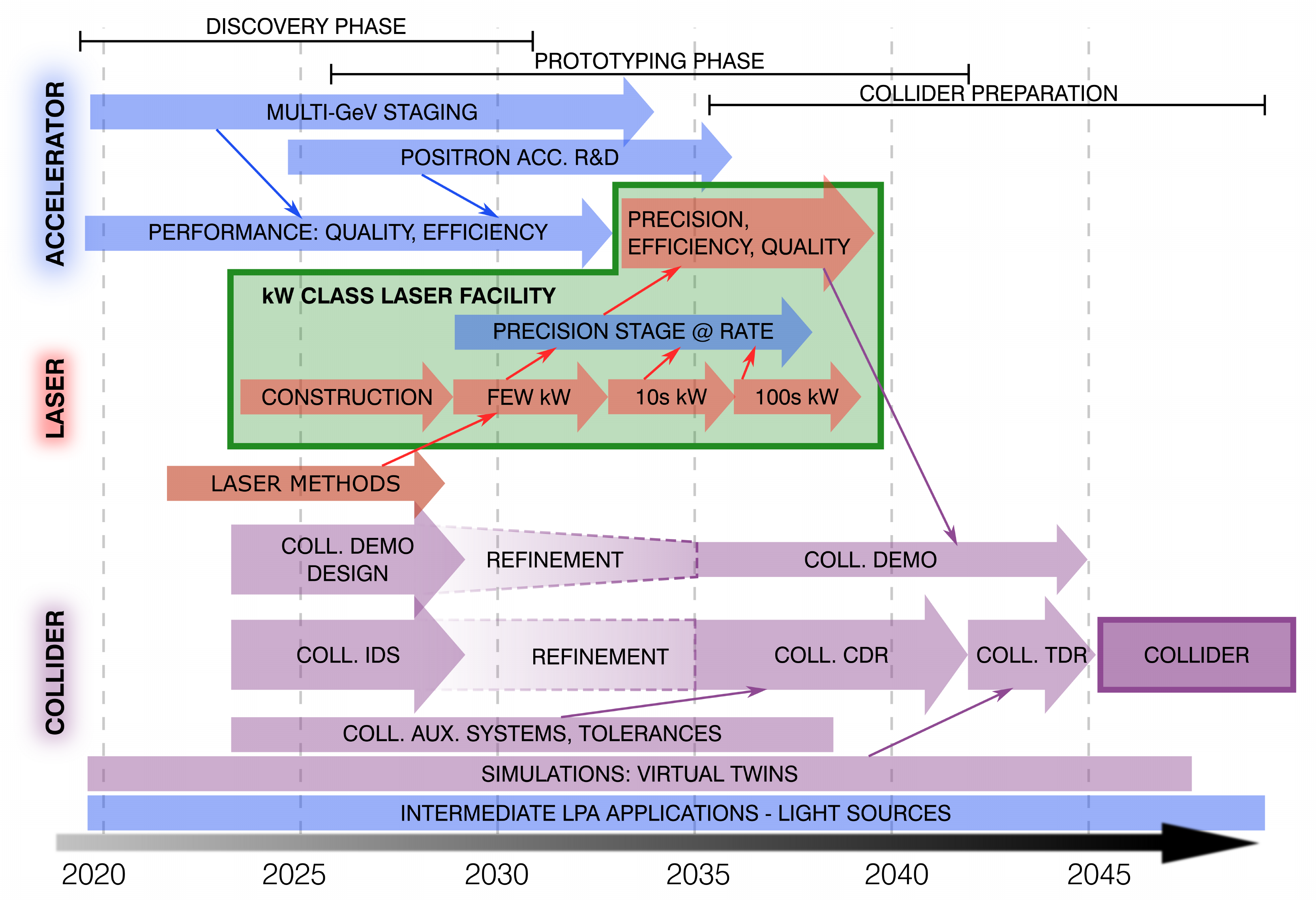}
\caption{\label{fig:new_road} Technically-limited high-level laser-plasma-accelerator-based collider R\&D roadmap.}
\end{figure*}

In 2016, a roadmap for R\&D toward a laser-plasma-based collider was identified and published in the DOE \emph{Advanced Accelerator Development Strategy Report} \cite{roadmap16}.  This report included R\&D for both laser-plasma acceleration technology and high peak and average power laser systems.  Significant progress on the roadmap has occurred over the past five years. However, many of the R\&D goals remain.  Here we propose an updated roadmap that reflects the advances that have occurred since the publication of the report in 2016.  Figure~\ref{fig:new_road} presents an updated R\&D timeline and goals.

Most research effort has focused on the plasma linacs.  The community has recognized the need to begin work on development of auxiliary collider systems (e.g., injector, positron source, damping, BDS, etc.) that are compatible with plasma accelerators, and take advantage of the unique characteristics of plasma accelerated beams. With the development and design of appropriate auxiliary collider systems, a self-consistent integrated design study (IDS) may be completed, as a precursor to a conceptual design report. It is recognized by the community that an IDS is a top priority to advance the technical readiness of an LPA collider.

\subsection{Near- and mid-term applications}

Development of ultra-compact laser-plasma-based accelerator technology will have many applications and societal benefits beyond high-energy physics.  It is also recognized by DOE HEP that realization of near- and mid-term applications of LPAs will help advance the  technology toward a collider \cite{roadmap16}.  A white paper highlighting the relevance of emerging applications driven by advanced accelerator concepts has been submitted to Snowmass21 \cite{near-term-Snowmass21}.  Community outreach  and discussions during the Snowmass process yielded a short list of relevant applications.  These applications are divided into light sources (free-electron lasers, betatron x-ray sources, Thomson and Compton scattered gamma-ray sources), particle sources (ions and electrons, with medical applications), and fundamental applications (low-energy particle physics experiments and astrophysics exploration). A summary of near-term applications from the white paper Ref.~\cite{near-term-Snowmass21} is shown below in Table~\ref{tab:my_label}.

\begin{table}
    \centering
    \scalebox{0.75}{
 \begin{tabular}{|p{0.3\linewidth}|>{\centering\arraybackslash}p{0.3\linewidth}|>{\centering\arraybackslash}p{0.3\linewidth}|>{\centering\arraybackslash}p{0.3\linewidth}|}
    \hline
         \textbf{Source} & \textbf{Example application} & \textbf{Status} & \textbf{Readiness in 5--10 years}\\ \hline
        Plasma-based FEL \cite{Emma2021} & Single-shot high-res imaging, non-linear excitation & Experimental feasibility demonstrated, two high-impact papers in 2021 & Realistic at higher flux and photon energy in $<$ 5 years  \\ \hline
        Betatron X-rays  \cite{Corde_RMP_2013,Albert_PPCF_2016} &  Single-shot phase-contrast imaging of micro-structures    & Extensive demonstrations     &  Ready now  \\ \hline
        Compton-scattered X-rays \cite{Corde_RMP_2013,Albert_PPCF_2016} & Compact dose-reduced medical imaging, HED dynamics &   Proof-of-principle demonstrations  & Tunable and mono-energetic in $<$ 5 years  \\ \hline
        Advanced gamma ray sources \cite{Glinec_PRL_2005,Benismail_NIMA_2011,Gadjev2019,Sudar2020} & Security, efficient imaging at reduced dose & Experimental demonstrations (plasma based) & Plasma-based ready now. \\ \hline
        VHEE \cite{Svendsen2021,Labate2020,Kokurewicz2019} & Low dose radiotherapy & Well established, needs stability emphasis &  Ready now at compact low rep rate sources \\ \hline
        Laser plasma ions  \cite{Linz2016,Tochitsky20}&  Medical imaging, FLASH therapy, HED diagnostic & Extensive demonstrations in TNSA regime & Ready now, $>$100 MeV protons in $<$5 years\\ \hline
    \end{tabular}}
    \caption{High-level summary of near-term applications of advanced accelerators described in Ref.~\cite{near-term-Snowmass21}. References, example applications, status and readiness in 5--10 years are included for each source.}
    \label{tab:my_label}
\end{table}

\section{Required Experimental Research Facilities}%

Intense, short-pulse, laser-plasma interactions are an area of very active research in the U.S.\ and internationally.  There are a number of high-power laser facilities in the U.S.\ researching compact laser-plasma accelerated electron beams for high-energy physics applications.   Some notable PW laser facilities \cite{Danson19} performing HEP-relevant research are the BELLA PW laser facility at LBNL (40~J, 30~fs, at 1~Hz), the Titan laser at LLNL (130~J, 0.7~ps, at 1 shot per hour), the Hercules laser at the University of Michigan (15~J, 30~fs, at 1 shot per min), the Texas PW laser at the University of Texas-Austin (155~J, 150~fs, at 1 shot per hour), the Diocles laser at University of Nebraska (20 J, 30~fs, at 0.1~Hz), and the ALEPH laser facility at CSU (26~J, 30~fs, at 3~Hz).  Several sub-PW laser systems in the U.S.\ are also doing significant and important HEP-relevant LPA research.  In addition, it is expected that multi-PW, short-pulse laser systems will become available for LPA R\&D in the next several years, such as the ZEUS Laser Facility (3~PW) at the University of Michigan \cite{zeus} and the proposed EP-OPAL laser facility at the University of Rochester \cite{Bromage_2021} that is being designed to produce 2$\times$25~PW beams (20~fs, 500~J, 5~min shot-cycle time) with two target areas supporting a variety of propagation geometries.

Although the peak laser power of sub-PW and PW laser systems is already sufficient to perform relevant LPA experiments, yielding multi-GeV electron beam energy gain, new and/or upgraded facilities are required to develop and advance high-average power LPA technology. 
Short-pulse laser technology is rapidly advancing and it is expected that Joule-class, kW average power short pulse systems will be available in the next several years \cite{SnowmassLaser21}. Developing LPAs to kHz repetition rates is critical as a step toward a collider.  In addition, many societal, nearer-term applications demand kHz rate.  Air and ground motion at of order 100~Hz has been shown to  generate fluctuations that limit the precision of current laser systems.  Hence, development of kHz laser-plasma accelerators with active feedback is required to advance laser-plasma accelerator technology and is expected to open a new era of precision timing and alignment, taking advantage of machine learning methods.  

A new or upgraded facility is required to develop kHz LPAs, including development high-rep rate targetry, diagnostics, and feedback techniques.  One possible facility to meet these needs is the proposed kBELLA (kHz BErkeley Lab Laser Accelerator) project to develop a kW average power, kHz laser-plasma accelerator facility to advance the research on laser-plasma-based acceleration.
Several recent community planning \cite{BLI18, DPP20}, Basic Research Needs \cite{BRN19}, and National Academies  \cite{BrightestLightReprot18,NAS-Decadal20} reports have highlighted the U.S.\ scientific community consensus that such a facility is required to advance  a wide range of science and applications as well as to maintain U.S.\ competitiveness given strong international investments in laser technology development. 
 
\section{Summary and Recommendations}

Laser-plasma accelerators (LPAs) are capable of sustaining accelerating fields of 10-100 GeV/m, 100-1000 times that of conventional RF technology, and  the highest fields produced by any of the widely researched advanced accelerator concepts. Furthermore, LPAs intrinsically produce short particle bunches, 100-1000 times shorter than that of conventional RF technology, which leads to reductions in beamstrahlung and, hence, savings in the overall power consumption to reach a desired luminosity.  Furthermore, they enable novel energy recovering methods that can reduce power consumption and improve the luminosity per unit energy consumption for linear colliders. These properties make LPAs a promising candidate as drivers for a more compact, less expensive high-energy collider by providing  multi-TeV polarized leptons in a relatively short distance $\sim$1~km. Collider concepts are discussed up to the 15~TeV range. A future RF-based linear collider facility could be re-purposed to deliver higher energies with LPA technology thereby extending physics reach while saving on construction costs.

Several previous reports have made recommendations for a vigorous program on LPA R{\&}D and applications, including the previous P5 and HEPAP subcommittee reports, the European Strategy and Laboratory Directors Group reports, and several others (see Sec.~\ref{sec:reports}). Numerous significant results have been obtained since the last P5 report, including the production of high quality electron bunches at 8~GeV from a single stage, the staging of two LPA modules, novel injection techniques for ultra-high beam brightness, investigation of processes that stabilize beam break up, new concepts for positron acceleration, and new technologies for high-average-power, high-efficiency lasers. In addition to the long term goal of a high energy collider, LPAs can provide compact sources of particles and photons for a wide variety of near-term applications in science, medicine, and industry.

Research on LPAs has exploded in recent years, driven in part by the extremely rapid advances made in high-power lasers based on the 2018 Nobel Prize winning technique of chirped-pulse amplification. Numerous high-power laser facilities have sprung up worldwide, particularly in Europe and Asia. Consequently, about 800--1000 research papers are published annually on LPAs (cf.~Fig.~\ref{fig:nrofpubl}). Since much of this research is overseas, it is critical that the U.S.\ make strong investments in LPAs to ensure global leadership.

The LPA community proposes the following recommendations to the Snowmass conveners:

\begin{enumerate}
\item {\bf  Vigorous research on LPAs in the GARD program}: We recommend  a vigorous research program on LPAs,  including experimental, theoretical, and computational components, continuing and enhancing the present effort in the GARD program. This will ensure rapid progress along the LPA R{\&}D roadmap towards a high energy collider and to develop intermediate applications. Topics include development of high-fidelity multi-GeV LPA modules, the high efficiency staging of modules, tailoring and control of laser and plasma profiles, laser injection techniques that produce ultra-high brightness beams, positron acceleration, and compact cooling and beam delivery systems. This support is also required to ensure international competitiveness. This, along with recommendations 2 and 3, will lay the ground work for the integrated design study in recommendation 4.

\item {\bf Enhanced R{\&}D on laser drivers}:  {R\&}D on laser technology is needed to develop the laser drivers that power LPA colliders. An LPA collider at high energy and high luminosity requires high power laser drivers at high rep-rates, tens of kHz,  and high average powers,  hundreds of kW,  with high efficiencies. New laser technologies are being developed that can deliver 1 kHz rep-rates in the near term and that scale to tens of kHz in the long term. Additional resources are needed beyond the present efforts supported by the DOE Accelerator Stewardship Program. 

\item {\bf  Near-term LPA capability extensions}:
Near-term upgrades and expansions to the existing LPA facilities are required to ensure rapid progress along the LPA roadmap and to remain competitive with international facilities. These include laser performance enhancements, pulse shaping techniques, additional beam lines, feedback and control systems, advanced diagnostics, as well as the construction of a new high rep-rate (kHz) laser user facility for precision LPAs to support R{\&}D towards collider performance requirements.

\item {\bf  Integrated design study of a high energy (1--15 TeV) LPA collider}:
An integrated design study of a high energy (1--15 TeV) LPA-based collider is needed to  detail all the components of the system, such the injector, drive laser, plasma source, beam cooling, and beam delivery system. This is in addition to the design of the LPA linacs, which has been the focus of the much the present LPA research. This study will include start-to-end simulations. Completion of the integrated design study would be followed by a conceptual design report.

\item {\bf Study of a collider demonstration facility at intermediate energy}: A design is needed for a collider demonstration facility at an intermediate energy (20--100 GeV). This would serve as a stepping stone facility on the roadmap to a high energy collider that would both demonstrate essential machine technologies and provide a user facility to study science at intermediate energies. This could consist of a combination of advanced accelerator technologies including laser and beam driven structures and plasmas. Possible configurations include a single linac composed of multiple stages for fix target studies, two electron linacs for a gamma-gamma collider, or an electron and positron linac for an intermediate energy lepton collider.   

\item {\bf DOE-HEP workshop on advanced accelerator strategy}:
A DOE-HEP workshop should be held in the near term to update and formalize the advanced accelerator strategy and roadmaps towards a high energy collider.

\end{enumerate}

\acknowledgments
This work was supported by the Director, Office of Science, Office of High Energy Physics, of the U.S. Department of Energy under Contract No.\ DE-AC02-05CH11231.


\bibliography{snowmass}

\end{document}